\documentstyle[preprint,aps,epsfig,axodraw]{revtex}

\def\PLBnew#1#2#3{{\em Phys. Lett.} {\bf B#1} (20#2) #3}
\def\PLB#1#2#3{{\em Phys. Lett.} {\bf B#1} (19#2) #3}

\def\PRD#1#2#3{{\em Phys. Rev.} {\bf D#1} (19#2) #3}
\def\PRL#1#2#3{{\em Phys. Rev. Lett.} {\bf#1} (19#2) #3}

\def\gev{\mbox{ GeV}}

\begin{document}
\newcommand{\cross}{\mbox{$\rlap{\kern0.125em/}$}}
\draft
\preprint{\vbox{\hbox{USM-TH-103}}}
\title{New Limits on the Color-Octet Technirho Mass from its Decay to Dijets}
\author{Alfonso R. Zerwekh }
\address{
 {\it Departamento de F{\'{\i}}sica,
Universidad T\'ecnica Federico Santa Mar{\'{\i}}a, \\ 
Casilla 110-V, Valpara{\'{\i}}so, Chile}
}
\maketitle
\widetext
 \begin{abstract}
  Recently has been shown that the physical (mass
  eigenstate) color-octet technirho does not couple to two gluons. In
  this letter we study how this result affects the presently accepted
  limits for the color-octet technirho mass obtained from the study of
   dijets production at the Tevatron. First we show that data from
  Tevatron Run 1b can not exclude any mass range. Finally, we obtain
  limits for the Tevatron Run II.  
\end{abstract}

\section{Introduction}

It is commonly believed that colliders such as the Tevatron (Run II)
and the LHC may discover phenomena beyond the Standard Model  and
reveal the mechanism of electroweak symmetry  breaking. Many people  believe
that this new physics will be described by some sort of super-symmetric
extension of the Standard Model. Nevertheless, other logical
possibilities still exist and must be considered. One of these
alternatives is dynamical electroweak symmetry breaking \cite{dsb}. Its best
known realization, the na{\"{\i}}ve QCD-like Technicolor idea, has been
severely challenged by precision measurement from LEP; however models
incorporating ideas such as 
Walking Technicolor or Top-color assisted Technicolor can evade these
difficulties. In this letter we deal with some features which are
present in many non-minimal Technicolor models.

 A very interesting prediction of
this kind of models is the existence of  color-nonsinglets
resonances. Of special interest is the color octet isosinglet vector
resonance called technirho. Its phenomenology has been studied by many
authors \cite{lr,eichten-lane} and  experimental limits on its mass
has been established at the Tevatron \cite{CDF1,CDF2}. However,
Zerwekh and Rosenfeld \cite{us}
have shown that the traditional description of the color-octet
technirho interactions can be considered incomplete. They propose a
complete and explicitly gauge invariant model for the color-octet
technirho interactions. An important consequence of their model is
that a physical technirho does not couple to two gluons. In this
letter we investigate how this result modifies the existing limit on
the technirho mass coming from its decay to dijets.

The letter is organized as follow. In Section \ref{sec:method} we present the
methodology we use in our calculation. Section \ref{sec:results} is
devoted to the presentation of our results. Finally, in Section
\ref{sec:conclusion}  we summarize our main conclusions.

\section{Methodology}
\label{sec:method}

Because the physical technirho does not couple to two gluons, its only
contribution to dijets production is through its
decay to a quark-anti-quark pair. The technirho-quarks interaction
is described by the lagrangian 
          
\begin{equation}
  {\cal L}_{\mbox{\small{int}}}= -g\tan\alpha \bar{\psi}\gamma^{\mu}\rho^a_{\mu}\frac{\lambda^a}{2}\psi 
\end{equation}
\noindent
where $g$ is the QCD coupling constant and $\alpha$ is the
technirho-gluon mixing angle with its value given by the formula:

\begin{equation}
  \sin \alpha =
  \frac{g}{\sqrt{2\pi}}\frac{1}{\sqrt{2.91\times 3/N_{tc}}} 
\end{equation}
\noindent
where $N_{tc}$ is the number of technicolors. In this letter we take
the usual value $N_{tc}=4$.

We wrote a FORTRAN code in order to convolute the partonic amplitudes
with the partonic distribution functions CTEQ2L \cite{cteq2}. In order
to compare 
our results with experimental data, the following kinematical cuts
were implemented \cite{CDF2}:

\begin{equation}
  \left|\eta\right| < 2.0 \gev \mbox{ for both jets}
\end{equation}
and

\begin{equation}
  \left|\cos \theta^*\right| = \left| \tanh\left[\frac{(\eta_1-\eta_2)}{2}\right]\right| < \frac{2}{3}  
\end{equation}
where $\eta$ is the pseudo-rapidity.

In our calculation we used the one loop expression for the running QCD
coupling $\alpha_{QCD}(Q^2)$ with $\Lambda_{QCD}=200$ MeV,  a
renormalization scale given by $Q^2=\hat{s}$\footnote{$\hat{s}$ is
  the partonic center of mass energy and in our case corresponds to the
  dijets invariant mass}and five
light quark flavors. We also assume that the technirho decay into two
technipions is closed.

\section{Results}
\label{sec:results}

Table \ref{tab:1} shows the technirho total width, for different values of the
technirho mass, and the signal cross section for a $p\bar{p}$ collider
running at a  center of mass energy given by $\sqrt{s}=1800$
GeV. 

In figure 1 we compare our result(continuous line) with the prediction obtained
using the traditional description of  the gluon-technirho mixing
(dashed line) and
with the  $95 \%$ C.L. upper limit obtained by CDF Collaboration
\cite{CDF2} for a luminosity 
of  $106$ pb$^{-1}$ (black triangles). An extrapolation of the
experimental upper limit for a luminosity of $2000$ pb$^{-1}$
(Tevatron Run II) can be obtained from the previous one  by scaling
inversely with the square root of the luminosity \cite{tev2000}, and
it is also plotted in the same figure (white triangles). 

Notice that our result is significantly smaller than the prediction of
the traditional approach. In fact,according to our result, the CDF
data cannot exclude any technirho mass range, which  is in contradiction with
the accepted limits: $260 < M_{\rho_{T8}} < 470$ GeV.

On the other hand, two mass intervals can be excluded at the Tevatron
Run II:  $260 < M_{\rho_{T8}} < 475$ GeV and $600 < M_{\rho_{T8}} <
775$ GeV, but an important intermediate mass interval is still
allowed. Again, our result  is in contradiction with the limits expected by
the traditional approach: $220 < M_{\rho_{T8}} < 820$ GeV.

\section{Conclusions}
\label{sec:conclusion}

In this letter we have studied the limits of the color-octet technirho
through its decay to dijets in the context of a recently proposed
complete model for the technirho 
interactions. Our results are significantly different from what is
expected in the framework of the traditional approach to the
technirho-gluon mixing. We show, for example, that the data from the
Tevatron Run Ib can not exclude any technirho mass range. On the other
hand, the Tevatron Run II can exclude masses in the range $260 <
M_{\rho_{T8}} < 475$ GeV and $600 < M_{\rho_{T8}} < 775$ GeV

\vspace{1.0 cm}

{\Large \bf Acknowledgments}

We would like to thank Claudio Dib and Rog\'erio Rosenfeld for their
critical reading of the paper.
This work was supported by a C\'atedra Presidencial 1997 (Iv\'an Schmidt).

\begin{table}[htbp]
  \begin{center}
    \caption{Technirho total width and cross section for the process
      $p\bar{p}\rightarrow \rho_{T8}\rightarrow \mbox{dijets}$ at
      $\sqrt{s}=1800$ GeV }
    \label{tab:1}

    \begin{tabular}{c|c|c||c|c|c}
$M_{\rho_{T8}}$(GeV)&$\Gamma_{\rho_{T8}}$(GeV)&
$\sigma$ (pb)& 
$M_{\rho_{T8}}$(GeV)&$\Gamma_{\rho_{T8}}$(GeV)&
$\sigma$ (pb)\\
\hline
200& 2.50& 264& 600& 8.92& 1.26\\
\hline
300& 3.75& 55.0&700&10.45&0.420\\
\hline
400& 5.67& 13.2&800&11.96&0.142\\
\hline
500& 7.36& 3.92&900&13.47&0.0501
    \end{tabular}
  \end{center}
\end{table}
   
\newpage

{\Large \bf Figure Captions}

\vspace{2cm}
{\bf Figure 1:} Our prediction for the
signal cross section(continuous line), the CDF upper limits for a
luminosity of 
106 pb$^{-1}$ (black triangles) and its extrapolation for a luminosity
of 2000 pb$^{-1}$ (white triangles). Also shown is the cross section
predicted by the traditional approach (dashed line)\\

\newpage
\begin{figure*}[t]
\begin{center}
\epsfig{file=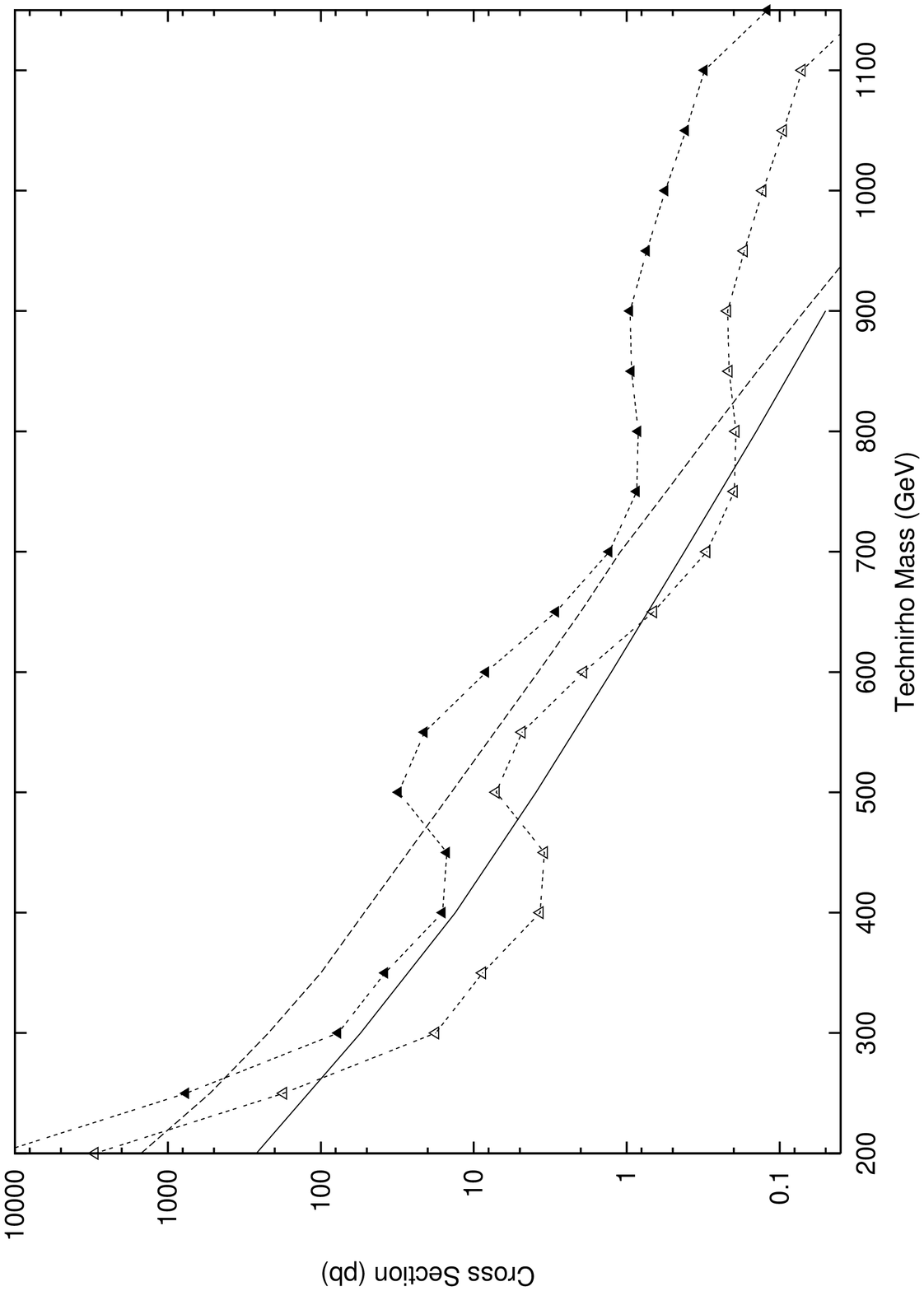,angle=0,width=15cm,height=20cm} \@
\end{center}
\end{figure*}

\end{document}